\documentclass[fleqn,twoside]{article}
\usepackage{espcrc2}

\usepackage{graphicx}
\usepackage{dcolumn}
\usepackage{bm}
\usepackage[english]{babel}
\usepackage[latin1]{inputenc} 
\usepackage{graphicx}
\usepackage{epstopdf}
\usepackage{amsfonts}
\usepackage{color}
\usepackage{epsfig}
\usepackage{mathrsfs}
\usepackage{amsmath,amssymb}
\usepackage{subfigure}

\usepackage[numbers, sort&compress]{natbib} 

\renewcommand{\(}{\left(}
\renewcommand{\)}{\right)}
\renewcommand{\[}{\left[}
\renewcommand{\]}{\right]}

\begin{document}

\graphicspath{{figure/}}
\selectlanguage{english}

\title{Anisotropy studies with multiscale autocorrelation function}

\author{M. De Domenico\address[SSC]{Laboratorio sui Sistemi Complessi, Scuola Superiore di Catania, Catania}\address[INFN]{Istituto Nazionale di Fisica Nucleare, Sez. di Catania, Catania (Italy)},
H. Lyberis\address[IPN]{CNRS/IN2P3 - IPN Orsay, Paris, France}
\address[UNITO]{Dipartimento di Fisica, Universit\'a di Torino, Torino, Italy}}

\begin{abstract}
We present a novel method, based on a multiscale approach, for detecting anisotropy signatures in the arrival direction distribution of the highest energy cosmic rays. This method is catalog independent, i.e. it does not
depend on the choice of a particular catalog of candidate sources, and it provides a good discrimination power even in presence of contaminating isotropic background. We present applications to simulated data sets of
events corresponding to plausible scenarios for events detected, in the last decades, by world-wide surface detector-based observatories for charged particles.
\vspace{1pc}
\end{abstract}


\maketitle


\section{Introduction}

In many field involving data analysis, the search for anisotropy has played a crucial role.
Many estimators, namely two-point correlation functions \cite{Peebles-1980,Davis-1983, Szalay-1993, Hamilton-1993}, have been proposed and widely used  to search for clustering of objects and to measure deviation from isotropy of angular distributions. A non-differential variant of such an estimators is widely used for small data set of points \cite{kachelriess2005ultra, kachelriess2006clustering, cuoco2006first, cuoco2008clustering, cuoco2009global}. These methods apply to angular coordinates of objects as well to distributions of arrival directions of events: in this work we will indifferently refer to both as \emph{arrival direction distributions of events}. 

Recently, new estimators have been introduced to study the anisotropy signature of sky's arrival direction distributions: the modified two-point Rayleigh \cite{ave20092pt+}, and shape-strength method derived from a principle component analysis of triplets of events \cite{Hague09}. Such a test statistics have been recently applied to both P. Auger data and to synthetic maps of events, the latter generated by sampling the Veron-Cetty \& Veron catalog \cite{VCV06} of nearby candidate active galactic nuclei (AGN) within 75 Mpc ($z\leq0.018$), showing an higher discrimination power than other estimators \cite{MelloNeto2009}.

Within the present work, we introduce a new fast and simple method for anisotropy analysis, which makes use of a multiscale approach and depends on one parameter only, namely the angular scale of the instrinsic anisotropy. The main advantage of our estimator is the possibility to analytically treat the results: the analytical approach drastically reduces computation time and makes available the possibility of applications to very large data sets of objects. We test the method on several simulated isotropic and anisotropic arrival direction distributions (mock maps) and perform an extensive analysis of its statistical features under both the null and the alternative hypotheses. However, it is worth remarking that the scope of applicability of our method is not limited to UHECR physics, and it is valid for any distribution of angular coordinates of objects.


\section{Multiscale Autocorrelation Function}\label{Scale}

Let $\mathcal{S}$ be a region of a spherical surface and let $P_{i}\(\phi,\theta\)$ ($i=1,2,...,n$) be a set of points locating $n$ arrival directions on $\mathcal{S}$, defining a \emph{sky}. The sky $\mathcal{S}$ is partitioned within a grid of $N$ equal-area (and almost-equal shape) disjoint boxes $\mathcal{B}_{k}$ ($k=1,2,...,N$) as described in Ref. \cite{stokes2004using}. Let $\Omega$ be the solid angle covered by $\mathcal{S}$, whereas each box $\mathcal{B}_{k}$ covers the solid angle 
\begin{eqnarray}
\Omega_{k} = \frac{1}{N}\int_{\theta_{\text{min}}}^{\theta_{\text{max}}}\int_{\phi_{\text{min}}}^{\phi_{\text{max}}}d\cos\theta d\phi = 2\pi(1-\cos\Theta)\nonumber
\end{eqnarray}
where $2\Theta$ is the apex angle of a cone covering the same solid angle: $N,\Theta$ and $\Omega$ are deeply related quantities that define a scale.

Let $\psi_{k}(\Theta)$ be the density of points in the data set that fall into the box $\mathcal{B}_{k}$: the function $A(\Theta)$ that quantifies the deviation of data from an isotropic distribution at the scale $\Theta$, is chosen to be the Kullback-Leibler divergence \cite{kullback1951information, kullback1987kullback}
\begin{eqnarray}
\label{def-A}
A(\Theta) =  \sum_{k=1}^{N} \psi_{k}(\Theta)\log\frac{\psi_{k}(\Theta)}{\overline{\psi}_{k}(\Theta)}
\end{eqnarray}
where $\overline{\psi}_{k}(\Theta)$, generally a function of the domain meshing, is the expected density of points isotropically distributed on $\mathcal{S}$ falling into the box $\mathcal{B}_{k}$. The Kullback-Leibler divergence is an information theoretic measure widely used in hypothesis testing and model selection criteria, statistical mechanics, quantum mechanics, medical and ecological studies (see \cite{dedomenico2010} and Ref. therein). This measure quantifies the error in selecting the density $\overline{\psi}(\Theta)$ to approximate the density $\psi(\Theta)$: it can be shown that minimizing the Kullback-Leibler divergence is equivalent to maximum likelihood estimation \cite{dedomenico2010}. It is straightforward to show that $A(\Theta)$ is minimum for an isotropic distribution of points, or, in general, when $\psi(\Theta)\sim \overline{\psi}(\Theta)$, i.e. if the model is correct.

If $A_{\text{data}}(\Theta)$ and $A_{\text{iso}}(\Theta)$ refer, respectively, to the data and to an isotropic realization with the same number of events, we define \emph{multiscale autocorrelation function} (MAF) the estimator
\begin{eqnarray}
\label{def-s}
s(\Theta)=\frac{\left|A_{\text{data}}(\Theta)-\left\langle A_{\text{iso}}(\Theta) \right\rangle\right|}{\sigma_{A_{\text{iso}}}(\Theta)}
\end{eqnarray}
where $\left\langle A_{\text{iso}}(\Theta) \right\rangle$ and $\sigma_{A_{\text{iso}}}(\Theta)$ are the sample mean and the sample standard deviation, respectively, estimated on several isotropic realizations of the data. If $\mathcal{H}_{0}$ denotes the null hypothesis of an underlying isotropic distribution for the data, the chance probability at the angular scale $\Theta$, properly penalized because of the scan on $\Theta$, is the probability 
\begin{eqnarray}
\label{def-p}
p(\Theta) = \text{Pr}\( s_{\text{iso}}(\Theta') \geq s_{\text{data}}(\Theta) | \mathcal{H}_{0}, \forall \Theta'\in\mathcal{P}\)
\end{eqnarray}
obtained from the fraction of null models giving a multiscale autocorrelation, at any angular scale $\Theta'$ in the parameter space $\mathcal{P}$, greater or equal than that of data at the scale $\Theta$. The null hypothesis is eventually rejected in favor of the alternative $\mathcal{H}_{1}=\lnot \mathcal{H}_{0}$ $-$ being $\lnot$ the negation operator $-$ at the angular scale $\Theta$, with probability $1-p(\Theta)$.

Under the null hypothesis $\mathcal{H}_{0}$, the estimator $s(\Theta)$ follows an half-gaussian distribution, independently on the value of the angular scale $\Theta$ and on the number of events on $\mathcal{S}$ \cite{dedomenico2010}.


\section{Dynamical Boxing}

The simplest definition of the boxing algorithm, as shortly described in Sec. \ref{Scale}, involves the fixed grid introduced in Ref. \cite{stokes2004using}, where each box only embodies the relative number of events falling in it. Unfortunately, such a \emph{static binning} approach could not reveal an existing cluster. Indeed, the fixed grid may cut a cluster of points within one or more edges, causing a further loss of information at the angular scale under investigation. To overcome this possible loss of information, we introduced the \emph{dynamical binning}, a type of smoothing of the grid by applying it on the data \cite{dedomenico2010}. 

The smoothing, adopted in our study, deals with a new counting procedure for the estimation of the density $\psi_{k}(\Theta)$. However, such a density depends on the observatory's exposure, generally a function on the celestial sphere depending on both the latitude of the experiment and the maximum zenith of detection, quantifying the effective time-integrated detection area for the flux of particles from each observable sky position. The relative exposure $\omega$ is the dimensionless function corresponding to the exposure normalized to its maximum value. For a single full-time operating detector, i.e. with constant exposure in right ascension, fully efficient for particles arriving with zenith angles smaller than $\theta_{\text{max}}$, it still depends on the declination $\delta$ \cite{sommers2001cosmic}. The proposed method takes into account the effect of non-uniform exposure on the arrival direction distribution of objects, by weighting the angular region around a given direction with the local value of the exposure. Our numerical studies show that such a \emph{dynamical binning} approach recovers the correct information on the amount of clustering in the data \cite{dedomenico2010}. 


\section{Interpretation of MAF}

Any catalog-independent method provides information about the angular scale $\Theta^{\star}$ where the significance is minimum. In the case of a simple two-point method, such an angular scale is quite difficult to interpret and topologically different configurations of events lead to the same significance. In the case of the modified two-point Rayleigh method, the estimation of the significance includes another set of parameters, independent from the angular distribution, as described in Ref. \cite{ave20092pt+}: parameters are sensitive to the orientation of the pairs and therefore to skies showing preferential directions and filamentary structure of points. It follows that $\Theta^{\star}$ is the most significant angular size for the mix of these informations, still linked to the pair configuration. In the case of the shape-strength method, the estimation of two parameters, namely the shape and strengh, is performed: both can be interpreted, respectively, in terms of size and elongation of the triangles defined by a triplet of points. It follows that all information is recovered from the configuration of triplets. 

\begin{figure}[!htb]
	\centering
	  \includegraphics[width=7.5cm]{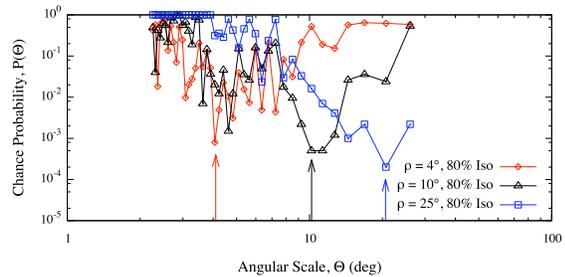}
	\caption{MAF: chance probability versus the angular scale for three mock skies of 60 events. In each sky, 20\% of events are normally distributed, with dispersion $\rho$, around a single source, and 80\% of events are isotropically distributed.}
\label{AngScaleMock}
\end{figure}

In the specific case of MAF, the angular scale $\Theta^{\star}$, where the significance is minimum, turns to be the significative \textit{clustering scale}: it is the scale at which occurs a greater accumulation of points respect to that one occuring by chance, with no regard for a particular configuration of points, e.g. doublets or triplets. Fig.\,\ref{AngScaleMock} shows the chance probability, as estimated by MAF, for three different point source mock skies contaminated by 80\% isotropic background. In each sky, 20\% of events are normally distributed, with dispersion $\rho$, around a single source: for the results shown in Fig.\,\ref{AngScaleMock}, we used three representative values for the dispersion, namely $\rho=4^{\circ}$, $\rho=10^{\circ}$ and $\rho=25^{\circ}$. Although the high isotropic background, the MAF method, by means of the dynamical binning, is able to recover the correct dispersion: in all cases, the most significant angular scale for clustering, indicated by arrows, recovers the dispersion of the corresponding sky.

\section{Statistical analysis of MAF}

In this section, we investigate the statistical features of MAF by inspecting its behavior under both the null or the alternative hypothesis. In particular, we estimate the significance $\alpha$ (or Type I error), i.e. the probability to wrongly reject the null hypothesis when it is actually true, and the power $1-\beta$ (where $\beta$ is known as Type II error), i.e. the probability to accept the alternative hypothesis when it is in fact true.

\vspace{0.25truecm}\hspace{0.1truecm}\emph{Null hypothesis.} We generate isotropic maps of $10^{5}$ skies, by varying the number of events from 20 to 500: for each sky in each map, we estimate the MAF for several values of the angular scale $\Theta$. Hence, we choose the value of $\Theta=\Theta^{\star}$ where the chance probability is minimum, as the most significant clustering scale: $\tilde{p}(\Theta^{\star})=\arg\min_{\Theta}p(\Theta)$, properly penalized because of the scan on the parameter $\Theta$, according to the definition in Eq. (\ref{def-p}). Indipendently on the number of events in the mock map, we find an excellent flat distribution of probabilities $\tilde{p}(\Theta^{\star})$, as expected for analyses under the null hypothesis $\mathcal{H}_{0}$, i.e. the MAF is not biased against $\mathcal{H}_{0}$.

Because of the definitions in Eq. (\ref{def-A}), (\ref{def-s}) and for the central limit theorem, an half-normal distribution is expected for the estimator $s(\Theta)$. We find an excellent agreement between the distribution for Montecarlo realizations and the expected one. It follows that the (unpenalized) probability to obtain by chance a value of the MAF, greater or equal than a given value $s_{0}$, is just $1-\text{erf}\(\frac{s_{0}}{\sqrt{2}}\)$, being erf the standard error function, independently on the angular scale $\Theta$ \cite{dedomenico2010}. Although this nice feature of the MAF estimator, generally the distribution of $s_{\text{max}}=\max\{s(\Theta)\}$ is of interest for applications, because of the required penalization due to the scan over the parameter $\Theta$. Hence, it is important to identify the distribution of the penalized probability $p(\Theta)$, if any. Intriguingly, our numerical studies show that such a distribution exists and it corresponds to one of the limiting densities in the extreme value theory (see \cite{dedomenico2010} and Ref. therein). In particular, the probability density of maxima is known as the Gumbel distribution \cite{gumbel1954statistical,gumbel2004statistics}:
\begin{eqnarray}
g(x)=\frac{1}{\sigma}\exp\[ -\frac{x-\mu}{\sigma} -\exp\(\frac{x-\mu}{\sigma}\) \]
\end{eqnarray}
where $\mu$ and $\sigma$ are the location and shape parameters, respectively.

\begin{figure}[!htb]
	\centering
	  \includegraphics[width=7.5cm]{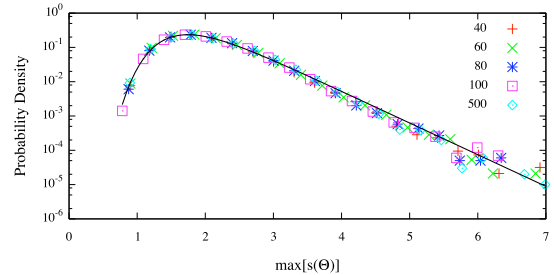}
	\caption{MAF: distributions of $\max\{s(\Theta)\}$ for $n=40, 60, 80, 100$ and $500$ events. Solid line correspond to the least-square fit of the Gumbel density with parameters $\mu=1.743  \pm 0.002$ and $\sigma=0.470   \pm 0.002$ ($\chi^{2}/\text{ndf}=1.1\times 10^{-5}$).}
\label{smax-gumbel}
\end{figure}

In Fig. \ref{smax-gumbel} are shown the probability densities of $s_{\text{max}}$ for $n=40, 60, 80, 100$ and $500$ events: independently on $n$, each density is in excellent agreement with the Gumbel distribution of extreme values, for the parameters $\mu=1.743  \pm 0.002$ and $\sigma=0.470   \pm 0.002$. Such a parameters correspond to the mean and to the standard deviation of the distribution, $\tilde{\mu}\approx 2.00$ and $\tilde{\sigma}\approx 0.59$, respectively (see \cite{dedomenico2010} and Ref. therein). It follows that the probability to obtain a maximum value of $s(\Theta)$, at any angular scale $\Theta$, greater or equal than a given value $\max\{s(\Theta)\}$ is
\begin{eqnarray}
p\(s_{\text{max}}\)=1-\exp\[ -\exp\(\frac{s_{\text{max}}-\mu}{\sigma}\) \],\nonumber
\end{eqnarray}
providing an analytical expression for the penalized probability defined in Eq. (\ref{def-p}).

\vspace{0.25truecm}\hspace{0.1truecm}\emph{Alternative hypothesis.} In order to investigate the behavior of MAF under the alternative hypothesis of an underlying anisotropic distribution of objects, we generate anisotropic maps of $10^{4}$ skies, by varying the number of events from 20 to 100. In general, the anisotropy of a sky depends on several factors: for instance, in the case of cosmic rays, it depends on the distribution of sources, on magnetic fields and on propagation effects as energy loss or the GZK cutoff \cite{nagano2000observations, bhattacharjee2000origin} (and Ref. therein). Thus, a more complicated approach is required for the Montecarlo realization of the maps. In order to estimate the power of MAF, we build reasonable anisotropic maps reflecting in part the real-world scenario, keeping in mind that our purpose is to build an anisotropic set of events for statistical analysis and not to generate events mimicking real data sets with the best available approximation. We proceed as follows: 
\begin{enumerate}
\item \emph{Catalog of candidate sources.} Although several models for production mechanisms of UHECR are available \cite{nagano2000observations, bhattacharjee2000origin} (and Ref. therein), \cite{hillas1984origin, hill1987ultra, berezinsky1997cosmic, berezinsky1997ultrahigh, venkatesan1997constraints, farrar1998correlation, fargion1999ultra, arons2003magnetars}, it is generally accepted that the candidate sources are extragalactic and trace the distribution of luminous matter on large scales \cite{waxman1997signature}. In particular, it has been shown that correlation with possible high redshift sources is unlikely \cite{sigl2001testing}, whereas compact sources are favored \cite{fodor2000ultrahigh, tinyakov2001correlation}: the recent result reported by the P. Auger Collaboration  experimentally supports the latter claim, showing an high correlation between the observed data and the distribution of nearby active galactic nuclei (AGN) \cite{auger2007correlation, auger2008correlation}. For these reasons, we use the Palermo Swift-BAT hard X-ray catalogue of AGN with known redshift within 200 Mpc ($z\leq0.047$) \cite{cusumano2010palermo}, as the reference catalog of candidate sources providing the most complete and uniform all-sky hard X-ray survey up to date.

\item \emph{Source effects.} Events, from each source in the reference catalog, are generated by weighting for the source flux and for the expected geometrical flux attenuation. Hence, the number of events coming from a source is proportional to its flux and to the factor $z^{-2}$: because of the small scales and the high energy of cosmic rays involved in anisotropy studies ($E\geq 4.0\times 10^{19}$ EeV), we assume a flat universe with zero cosmological constant ($\Omega=1$, $\Lambda=0$) and nonevolving source. Indeed, we naively take into account the possible deflections of the particles, due to the random component of the magnetic field, by producing arrival directions gaussianly-distributed with dispersion $\rho$ around the source. It is worth remarking that such a dispersion is strictly related to both the injection energy and the mass of the particle, as well as other physical quantities \cite{nagano2000observations}.

\item \emph{Background.} We take into account the possibility for a contaminating isotropic background of the anisotropy signal, by generating a number of events isotropically distributed, corresponding to a fraction $f_{\text{iso}}$ of the whole data set.

\item \emph{Detection effects.} As previously explained, the number of events detected by a single fully efficient and full-time operating surface detector, depends on its own relative exposure. In order to take into account such a detection effect, we generate the events according to the relative exposure of the single detector.
\end{enumerate}

However, for a more realistic distribution of events, several more constraints, in general based on further assumptions or debated models, are required: the mass of the particle, the injection spectrum of the source, the intervening magnetic field, to cite some of the most important. In our study of the MAF discrimination power, we fix $\rho=3^{\circ}$, as the mean angular deviation of UHECR in the galactic and extra-galactic magnetic field, and a background fraction $f_{\text{iso}}=0.3$.

\begin{figure}[!htb]
	\centering
	  \includegraphics[width=7.5cm]{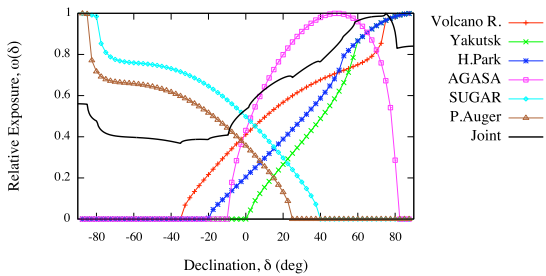}
	\caption{Relative geometrical exposure of each single detector listed in Tab. I (lines and points), and the joint exposure of all experiments (solid line).}
\label{all-expo}
\end{figure}

\begin{figure*}[!htb]  
  	\subfigure[\, UHECR events and candidate sources.]{ 
	  \includegraphics[width=8.5cm]{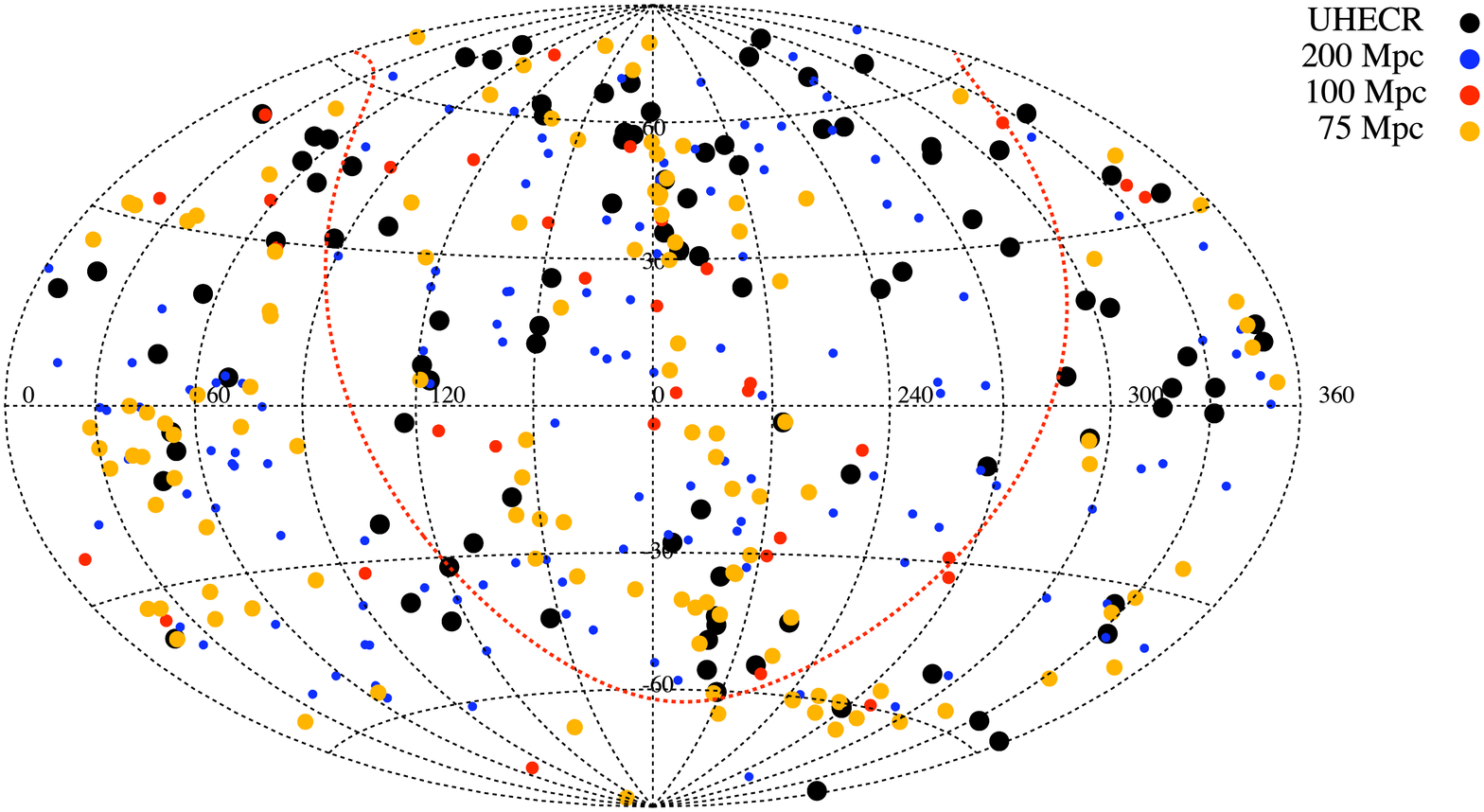}}
	\subfigure[\, Mock map.]{ 
	  \includegraphics[width=8.5cm]{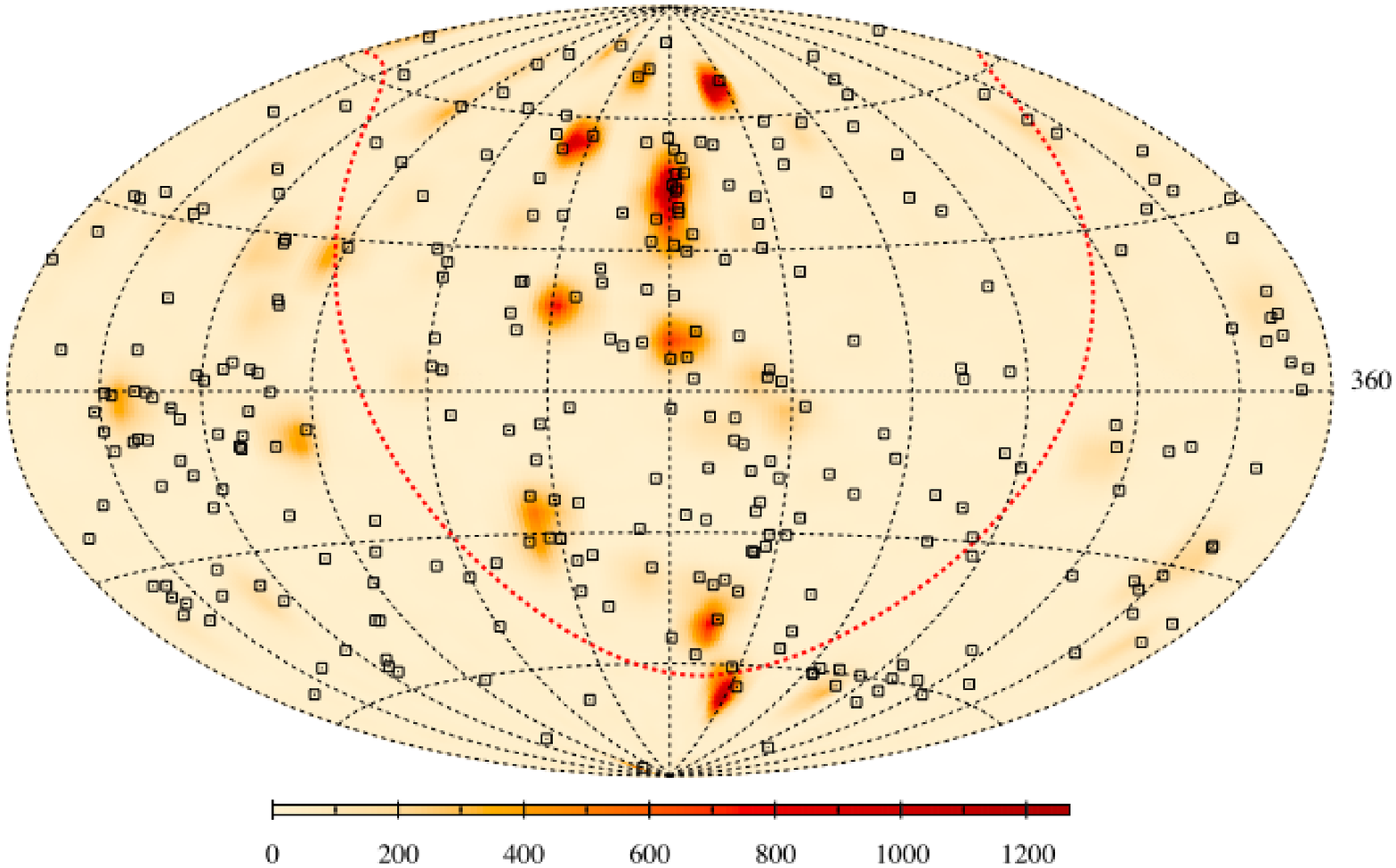}}
	  \caption{a) All-sky data set of 102 detected events with rescaled energy $E'\geq 40$ EeV (see the text for further information) superimposed on the distribution of AGN with known redshift ($z<0.047$) from the Palermo SWIFT-BAT hard X-ray catalogue; b) corresponding sources (squares) and smoothed mock map, generated for the statistical analysis (colour indicates the number of events). Equatorial coordinates are shown.}
\label{sky-data}
\end{figure*}

In order to produce a likely map of UHECR, we choose to generate events distributed in the whole sky, according to the number of events collected by surface detectors in the last decades. In particular, we consider events with energy $E\geq 4.0\times10^{19}$ EeV and error on the arrival direction smaller than $5^{\circ}$, as detected at the Sidney University Giant Airshower Recorder (SUGAR), Akeno Giant Air Shower Array (AGASA), Haverah Park, Volcano Ranch, Yakutsk and P. Auger Observatory (we refer to \cite{dedomenico2010} for a comprehensive list of references for the data released by each experiment). However, the fluxes of particles as measured by those experiments do not agree each other in the absolute fluxes, and a rescaling is needed \cite{berezinsky2009ultra}. By assuming that the spectrum reported by the HiRes Collaboration \cite{abbasi2008first} corresponds to the correct energy scale, the rescaling, based on some specific characteristics of the UHECR spectrum, fixes the energy shift factors $\lambda$ for the other experiments \cite{berezinsky2009ultra, kachelriess2006clustering}. Positions, maximum zenith angles $\theta_{\text{max}}$, exposures and energy shift factors are reported in Tab. I, for each experiment, as well as the number of detected events with rescaled energy $E'\geq 4.0\times10^{19}$ EeV ($E'=\lambda E$). In Fig. \ref{all-expo} is shown the relative geometrical exposure of each single detector listed in Tab. I, as well as the joint exposure of all experiments. For reference, in Fig. \ref{sky-data}a is shown the all-sky data set of 102 detected events with rescaled energy $E'$, superimposed on the distribution of AGN within 200 Mpc from the reference catalog, whereas in Fig. \ref{sky-data}b is shown the mock map of simulated events according to physical constraints previously described.

\begin{table}[!h]
\centering
\small
\begin{tabular}{lclcc}
\hline
\hline
\textbf{Experiment} & $\phi_{0}$ & $\theta_{\text{max}}$ & $\lambda$ & \#\textbf{Ev.}\\
\hline
\hline
Volcano R. & $35.15^{\circ}$N  &  $70^{\circ}$ & 1.000 & 6 \\
Yakutsk & $61.60^{\circ}$N &  $60^{\circ}$       & 0.625 & 20   \\
H. Park & $53.97^{\circ}$N &  $74^{\circ}$       & 1.000 & 7    \\
AGASA   & $35.78^{\circ}$N &  $45^{\circ}$     & 0.750 & 29   \\
SUGAR   & $30.43^{\circ}$S & $70^{\circ}$      & 0.500 & 13   \\
P. Auger & $35.20^{\circ}$S &  $60^{\circ}$      & 1.200 & 27   \\
\hline
\hline
\end{tabular}
\caption{Surface detectors: positions, maximum zenith angles $\theta_{\text{max}}$, exposures, energy shift factors and number of detected events with rescaled energy $E'\geq 4.0\times10^{19}$ EeV.}
\end{table}

\begin{figure}[!h]
	\centering
	  \includegraphics[width=7.5cm]{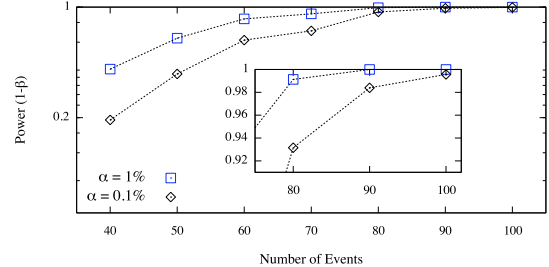}
	\caption{MAF power vs. the number of events sampled from anisotropic mock maps generated as described in the text, for values of the significance corresponding to $\alpha=0.1\%$ and $\alpha=1\%$.}
\label{maf-power}
\end{figure}

In Fig. \ref{maf-power} we show the power $1-\beta$ vs. the number of anisotropically distributed events, generated as described above. A sky is labelled as \emph{anisotropic} if, for a fixed value of the significance $\alpha$, the penalized chance probability as defined in Eq. (\ref{def-p}) is lesser or equal than $\alpha$, i.e. if the condition $\tilde{p}(\Theta^{\star})=\arg\min_{\Theta} p(\Theta)\leq \alpha$ holds for some angular scale $\Theta^{\star}$. In Fig. \ref{maf-power} is shown the power for two values of the significance threshold, namely $\alpha=0.1\%$ and $\alpha=1\%$, estimated through the analytical approach. For applications, a power of 90\% is generally required: under this threshold the method could miss to detect an existing anisotropy signal. In the case of the MAF, and for the considered anisotropic mock map, the power increases with the number of events $n$ and it is able to detect the anisotropic signal for $n\geq 60$, with significance $\alpha=1\%$. However, by decreasing the significance for the statistical test, the power requires a greater number of events to reach the 90\% threshold, as expected: our test clearly shows that the MAF provides an excellent discrimination power for $n\geq80$. Indeed, we verified the agreement between the analytical and the Montecarlo estimations of the discrimination power.



\section{Discussion and conclusion}

We introduced a new statistical test, based on a multiscale approach, for detecting an anisotropy signal in the arrival direction distribution of UHECR, that makes use of an information theoretical measure of similarity, namely the Kullback-Leibler divergence, and of the extreme value theory. Within the present work we showed that our procedure is suitable for the analysis of both small and large data sets of events, by applying it on several Montecarlo realizations of isotropic and anisotropic synthetic data sets, corresponding to plausible scenarios in the physics of highest energy cosmic rays. In fact, for small data sets as well as for larger ones, the method is able to recover the information about the most significant angular scale of clustering in the data, even in presence of strong isotropic contamination. 

The advantages of our approach over other methods are multiples. First, the method allows an analytical description of quantities involved in the estimation of the amount of anisotropy signal in the data, avoiding thousands of Montecarlo realizations needed for the penalizing procedure of results and drastically reducing the computation time. Second, the method allows the detection of a physical observable, namely the clustering scale, in the case of a point source. In the case of multiple sources, the information is about the most significant clustering scale(s), according to source distribution. Third, the method is unbiased against the null hypothesis and it provides an high discrimination power even in presence of strong contaminating isotropic background, for both small and large data sets. Although in this work we referred to UHECR physics for our applications, it is worth remarking that the method is suitable for the detection of the anisotropy signal in each data set involving a distribution of angular coordinates on the sphere, and it can be adapted to non-spherical spaces by properly redefining the dynamical boxing algorithm.

\textbf{Acknowledgments.} Authors thank the P. Auger Collaboration for generous comments and fruitful discussions and the ``Fondo per il potenziamento per la ricerca in informatica'', (Dipartimento di Matematica e Informatica, Università degli Studi di Palermo) for having kindly made available the computing resources required for the analyses presented within the present work.


\addcontentsline{toc}{section}{References} 
\begin{small}
\bibliographystyle{mprsty} 
\bibliography{draft}
\end{small}

\end{document}